\documentclass[a4paper,fleqn]{cas-dc}

\usepackage[english]{babel}
\usepackage[numbers,sort&compress]{natbib}
\usepackage{graphicx}
\usepackage{graphics}
\usepackage{braket}
\usepackage{bbold}
\usepackage{amssymb}
\usepackage{amsmath}
\usepackage{nicefrac}
\usepackage{dcolumn}% Align table columns on decimal point
\usepackage{bm}% bold math
\usepackage{slashed}
\usepackage{datetime}
\usepackage{mciteplus}
\usepackage{multirow}
\usepackage{bbold}
\usepackage{color}
\usepackage{xcolor}
\usepackage{float}
\usepackage[utf8]{inputenc}
\usepackage[normalem]{ulem}

\def\tsc#1{\csdef{#1}{\textsc{\lowercase{#1}}\xspace}}
\tsc{WGM}
\tsc{QE}
\newcommand{\tcite}[1]{~\cite{#1}}

\begin{document}
\let\WriteBookmarks\relax
\def\floatpagepagefraction{1}
\def\textpagefraction{.001}

\shorttitle{A journey into the proton structure: Progresses and challenges}    

\shortauthors{Celiberto, Francesco Giovanni}  

\title []{\Huge A journey into the proton structure \\ Progresses and challenges}  

\author[1,2,3]{Francesco Giovanni Celiberto}[orcid=0000-0003-3299-2203]

\cormark[1]

%\fnmark[<footnote mark no>]

\ead{fceliberto@ectstar.eu}

%\ead[url]{<URL>}

\affiliation[1]{organization={European Centre for Theoretical Studies in Nuclear Physics and Related Areas (ECT*)},
            addressline={Strada delle Tabarelle 286}, 
            city={Villazzano},
%          citysep={}, % Uncomment if no comma needed between city and postcode
            postcode={I-38123}, 
            state={Trento},
            country={Italy}}

\affiliation[2]{organization={Fondazione Bruno Kessler (FBK)},
            addressline={Via Sommarive 18}, 
            city={Povo},
%          citysep={}, % Uncomment if no comma needed between city and postcode
            postcode={I-38123}, 
            state={Trento},
            country={Italy}}

\affiliation[3]{organization={INFN-TIFPA Trento Institute of Fundamental Physics and Applications},
            addressline={Via Sommarive 14}, 
            city={Povo},
%          citysep={}, % Uncomment if no comma needed between city and postcode
            postcode={I-38123}, 
            state={Trento},
            country={Italy}}

%\cortext[1]{Corresponding author}



%-----------------------------------------
\begin{abstract}
 Unraveling the inner dynamics of gluons and quarks inside nucleons is a primary target of studies at new-generation colliding machines. Finding an answer to fundamental problems of Quantum ChromoDynamics, such as the origin of nucleon mass and spin, strongly depends on our ability of reconstructing the 3D motion of partons inside the parent hadrons.
 We present progresses and challenges in the extraction of TMD parton densities, with particular attention to the ones describing polarization states of gluons, which still represent a largely unexplored field. Then, we highlight connections with corresponding parton densities in the high-energy limit, the so-called unintegrated gluon distributions or UGDs and, more in general, to recent developments in high-energy physics.
\end{abstract}
%-----------------------------------------


%\item 
%\item 
%\item 

% Keywords
% Each keyword is separated by \sep

\maketitle

%-----------------------------------------
\section{Unveiling the multidimensional structure of the proton}
\label{sec:introduction}
%-----------------------------------------

The advent of new-generation particle accelerators\tcite{Begel:2022kwp,Dawson:2022zbb,Bose:2022obr,Anchordoqui:2021ghd,Feng:2022inv,Hentschinski:2022xnd,Acosta:2022ejc,AlexanderAryshev:2022pkx,Brunner:2022usy,Arbuzov:2020cqg,Abazov:2021hku,Bernardi:2022hny,Amoroso:2022eow,Celiberto:2018hdy,Klein:2020nvu,2064676,Black:2022cth,MuonCollider:2022xlm,Aime:2022flm,MuonCollider:2022ded}, such as the High-Luminosity Large Hadron Collider (HL-LHC)\tcite{Chapon:2020heu} and the Electron-Ion Collider (EIC)~\cite{Accardi:2012qut,AbdulKhalek:2021gbh,Khalek:2022bzd}, will start a new era in the search for long-waited signals of New Physics beyond the Standard Model (SM) of particle physics. At the same time, it has opened up a great window of opportunities for precision studies of the dynamics of \emph{strong interactions}, responsible for the strong nuclear force, that ``bind us all''. Among these analyses, deepening our knowledge of the proton content represents a primary target of efforts made by a very active scientific community, whose richness and spread grow over time.
The most powerful tool to gather information about the inner structure of the proton is the so-called language of parton correlators, that allows us to address the distribution of proton elementary constituents (quark and gluons, generally called \emph{partons}) in space, momentum and energy. Of special interest are the \emph{transverse-momentum-dependent} (TMD) parton distribution functions, which permit us to unravel the proton content in a three-dimensional representation, namely via a faithful \emph{3D tomographic imaging}. A TMD-like description is needed to find an answer to more fundamental questions, such as understanding the origin of the proton spin\tcite{EuropeanMuon:1987isl}.

From a physical viewpoint, TMD PDFs (or simply TMDs) come out as 3D generalizations of 1D collinear PDFs, since they account for struck-parton transverse-mo\-men\-tum effects.
A TMD description is required in order to correctly depict observables sensitive to the intrinsic transverse motion of colliding partons.
The latter,
quantified by TMDs, leaves its imprint on transverse momenta of particles tagged in the final state. It can be accessed through experimental measurements of semi-inclusive final states featuring observed transverse momenta or transverse imbalances much smaller than the reference hard scale(s).
To achieve a complete tomographic reconstruction of the proton, the information encoded in TMDs needs to be complemented by the one carried by another family of 3D maps: the \emph{generalized parton distributions} (GPDs).
While TMDs the give us access to the proton 3D content in momentum space, GPDs permit us to unveil the 3D dynamics of partons in the position space.
More in particular, information on the transverse position of a parton in the parent hadron is gathered from exclusive reactions with the
proton being deflected at small angles. 
GPDs quantify the parton transverse-spatial distribution via
a Fourier transform of the transverse momentum
transferred to the proton.

Formally, both TMDs and GPDs are projections of more general distributions, defined in terms of the fully unintegrated and off-diagonal correlator.
These quantities are known as \emph{generalized} TMDs (GTMDs) and they are often referred as ``mother'' distributions of TMDs and GPDs~\cite{Ji:2003ak,Belitsky:2003nz,Meissner:2008ay,Meissner:2009ww,Lorce:2011dv,Lorce:2011kd,Lorce:2011ni,Lorce:2013pza,Burkardt:2015qoa,Bertone:2022awq,Echevarria:2022ztg}.
However, beyond this purely formal link, no general relations between TMDs and GPDs are known. 
Some nontrivial connections have been found only via models\tcite{Burkardt:2002ks,Burkardt:2003uw,Meissner:2007rx}. 
Striking examples are the so-called \emph{chromodynamic lensing} relations\tcite{Burkardt:2002ks,Burkardt:2003je,Gamberg:2009uk,Pasquini:2019evu} connecting
time-reversal effects in single-spin asymmetries and spatial distortions of the GPDs in the impact-parameter
space (see also Ref.\tcite{Diehl:2015uka}).
These functional relations do not hold anymore when higher Fock states and/or higher-order diagrams are considered in models\tcite{Meissner:2007rx,Burkardt:2007xm,Bacchetta:2011gx,Pasquini:2007xz}.

%-----------------------------------------
\section{TMD gluon distribution functions}
\label{sec:gluon_TMDs}
%-----------------------------------------

Unpolarized and polarized gluon TMDs at twist-2 (leading twist) for a spin-1/2 target were identified for the first time in Refs.~\cite{Mulders:2000sh} (see also Refs.~\cite{Meissner:2007rx,Lorce:2013pza}).
TMD distributions are sensitive to the \emph{resummation} of logarithmically enhanced terms proportional to the observed transverse momentum. While our knowledge of this perturbative contribution is well known~\cite{Bozzi:2003jy,Catani:2010pd,Echevarria:2015uaa}, the genuine nonperturbative TMD content is a largely uncharted territory.
The distribution of linearly polarized gluons in an unpolarized nucleon, $h_1^{\perp g}$, is a key ingredient to explain spin effects observed in unpolarized-hadron collisions~\cite{Boer:2010zf,Sun:2011iw,Boer:2011kf,Pisano:2013cya,Dunnen:2014eta,Lansberg:2017tlc}, whose size is expected to grows in the low-$x$ domain\tcite{Celiberto:2021zww,Bolognino:2021niq}.
The Sivers function, $f_{1T}^{\perp g}$, tells us about the distribution of unpolarized gluons in a transversely polarized nucleon. It plays a core role to explain transverse-spin asymmetries rising in collisions of polarized hadrons.
Remarkably, the Sivers function can be accessed via unpolarized electron-nucleon scatterings thanks to its connection with the Odderon in the forward-QCD limit~\cite{Boussarie:2019vmk}.

\begin{figure*}[t]
 \centering
 \includegraphics[scale=0.25,clip]{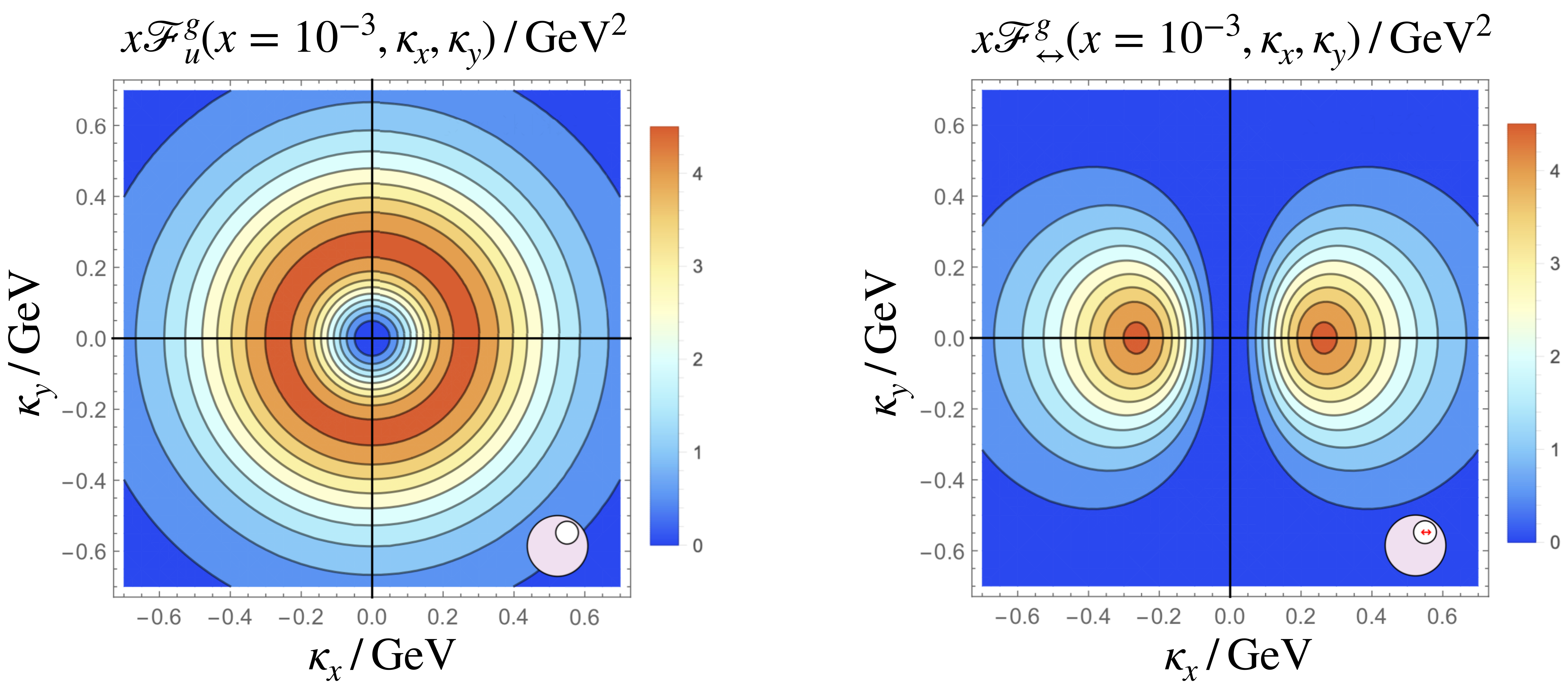}

 \caption{
 3D tomographic reconstruction of the unpolarized (left) and Boer--Mulders (right) gluon TMD distributions in the proton, as functions of the gluon transverse momentum, at the initial energy scale, $\mu_0 = 1.64$ GeV, and for $x = 10^{-3}$.
 Plots refer to replica 11.
 }
 \label{fig:3D_gluon_TMDs}
\end{figure*}

\begin{figure*}[t]
 \centering
 \includegraphics[scale=0.25,clip]{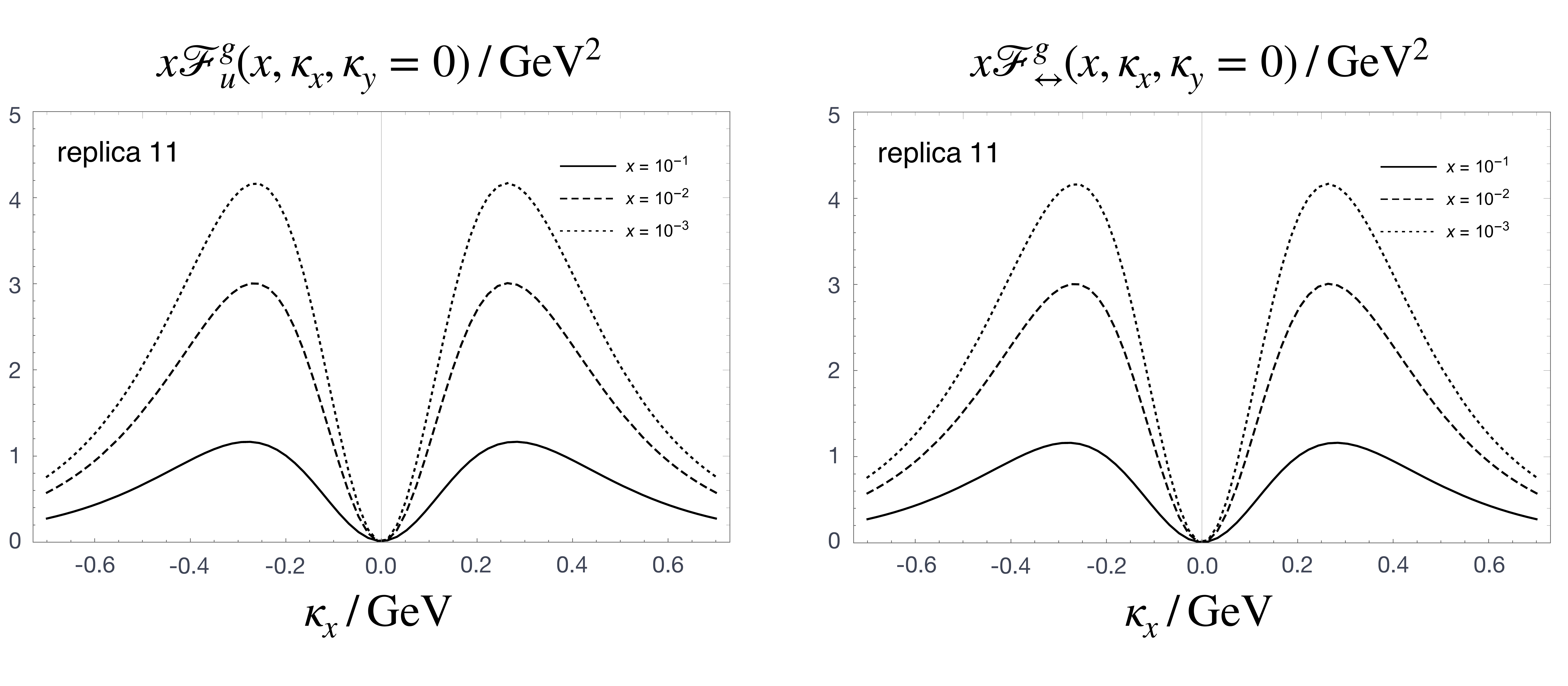}

 \caption{
 Ancillary, slice plots for the unpolarized (left) and Boer--Mulders (right) gluon TMD distributions in the proton, as functions of $\kappa_x$, while $\kappa_y = 0$, at the initial energy scale, $\mu_0 = 1.64$ GeV, and for $x$ ranging from $10^{-1}$ down to $10^{-3}$.
 }
 \label{fig:ancillary_gluon_TMDs}
\end{figure*}

TMDs turn out to be process-dependent due to the presence of transverse \emph{gauge links} (\emph{Wilson lines})~\cite{Brodsky:2002cx,Collins:2002kn,Ji:2002aa}.
Whereas quark TMDs depend on $[\pm]$ staple links that determine the direction of future- or past-pointing Wilson lines, gluon TMDs exhibit a more involved gauge-link dependence, since they are sensitive to combinations of two or more staple links. As a consequence, they feature a more complicated level of \emph{modified universality}.

There exist two principal gluon gauge links, known as $f\text{-type}$ and the $d\text{-type}$ structures. In the context of low-$x$ investigations they are often called Weisz\"acker--Williams and dipole links, respectively.
The antisymmetric $f_{abc}$ QCD color structure is typical of the $f$-type time-reversal odd (T-odd) gluon-TMD correlator, while the symmetric $d_{abc}$ structure determines the $d$-type T-odd one. This makes $f$-type ($d$-type) gluon TMDs dependent on $[\pm,\pm]$ ($[\pm,\mp]$) gauge-link combinations.
Much more intricate, box-loop structures rise along with reactions in which multiple color exchanges connect both the initial and final state~\cite{Bomhof:2006dp}. This fact, however, bring to violations of the TMD factorization~\cite{Rogers:2010dm,Rogers:2013zha}).

More in general, a rigorous proof of the validity of TMD factorization relies on establishing proper factorization theorems affording us a precise definition of initial-scale and evolved densities, thus defining their universality properties and evolution equations.
These theorems needs to be proven process by process.
Up to now, TMD theorems have been obtained for a few reactions sensitive to quark TMDs: Drell--Yan and semi-inclusive deep inelastic scattering at low observed transverse momenta\tcite{Collins:1981uk,Collins:2011zzd}, as well as lepton  annihilations into two almost back-to-back hadrons\tcite{Collins:1981uk,Collins:2011zzd,Bacchetta:2015ora}.
The factorization for the inclusive production of single hadrons has been discussed only recently\tcite{Kang:2020yqw,Makris:2020ltr,Boglione:2021wov}.

Conversely, proper factorization theorems for gluon-TMD sensitive processes are almost completely lacking, so that factorization is usually assumed in phenomenological studies. TMD factorization is expected to hold for gluon-induced color-singlet final states, such as Higgs-boson hadroproduction\tcite{Boer:2011kf,Boer:2013fca,Gutierrez-Reyes:2019rug}.
Quite recently, a factorization formula was derived for the inclusive hadroproduction of pseudoscalar quarkonium states\tcite{Echevarria:2019ynx} in the color-singlet channel\tcite{Einhorn:1975ua,Chang:1979nn,Berger:1980ni,Baier:1981uk}.
This led to an enhancement of the description of the quarkonium production mechanism that now includes TMD-like inputs, known as \emph{shape functions}\tcite{Echevarria:2019ynx,Fleming:2019pzj,DAlesio:2021yws}.
However, the bound-state nature of a quarkonium particle, instead of a point-like object, makes it difficult to extend that result to gluon-TMD studies through other quarkonia, such as vectors\tcite{Fleming:2019pzj,Boer:2020bbd,Lansberg:2017dzg,Scarpa:2019fol,Echevarria:2020qjk}.

From a phenomenological perspective, the gluon-TMD field is an almost uncharted territory. 
First studies of the unpolarized and the Sivers TMDs were performed Refs.~\cite{Lansberg:2017dzg,Gutierrez-Reyes:2019rug,Scarpa:2019fol} and~\cite{Adolph:2017pgv, DAlesio:2017rzj,DAlesio:2018rnv,DAlesio:2019qpk}, respectively.
Thus, exploratory analyses of gluon TMDs by means of flexible for the nonperturbative content of these densities are required. Progresses along this direction were proposed in the \emph{spectator-model} approach~\cite{Lu:2016vqu,Mulders:2000sh,Pereira-Resina-Rodrigues:2001eda}, which was formerly adopted to model quark TMDs~\cite{Bacchetta:2008af,Bacchetta:2010si,Gamberg:2005ip,Gamberg:2007wm,Jakob:1997wg,Meissner:2007rx}. In the gluon-TMD case, it assumes that the struck nucleon with mass $m_{\cal H}$ and momentum $\kappa_{\cal H}$ emits a gluon with momentum $\kappa$, transverse momentum $\boldsymbol{\kappa}_T$ and longitudinal fraction $x$. The remainders are effectively modeled by and effective on-shell particle with mass $m_{\cal X}$ and spin 1/2.
Within this framework, taken and at the diagrammatic tree level, all the initial energy-scale twist-2 TMDs can be calculated. Spectator-model gluon T-even functions were recently resented in Refs.~\cite{Bacchetta:2020vty,Celiberto:2021zww,Bacchetta:2021oht}.

In particular, the nucleon-gluon-spectator vertex was modeled as 
\begin{equation}
 \label{eq:form_factor}
 \Upsilon_{ab}^{\, \mu} = \delta_{ab} \left[ \eta_1(p^2) \, \gamma^{\, \mu} + \eta_2(p^2) \, \frac{i}{2 m_{\cal H}} \sigma^{\, \mu\nu}\kappa_\nu \right] \,,
\end{equation}
where the $\eta_{1,2}$ form factors are two dipolar functions of $\boldsymbol{\kappa}_T^2$.
In our model the spectator mass is not a fixed parameter, but it ranges from $m_{\cal H}$ to $+\infty$ and it is weighted through a spectral function ${\Xi}_{\cal X}^{[\rm s.m.]} (m_{\cal H})$, which reproduce both the small- and the moderate-$x$ behavior of gluon collinear PDFs.
The analytic expression of the spectral function contains seven parameters and reads
\begin{equation}
\label{eq:Xi_X}
 \Xi_{\cal X}^{[\rm s.m.]} (m_{\cal H}) = \mu^{2a} \left( \frac{\cal A}{{\cal B} + \mu^{2b}} + \frac{\cal C}{\pi \sigma} e^{-\frac{(m_{\cal H} - m_{\cal D})^2}{\sigma^2}} \right) \,.
\end{equation}

In Ref.\tcite{Bacchetta:2020vty} model parameters were fixed by making a simultaneous fit of unpolarized and helicity TMDs, $f_1^g$ and $g_1^g$, to their collinear counterparts from {\tt NNPDF}\tcite{Ball:2017otu,Nocera:2014gqa} at the initial scale of $\mu_0 = 1.64$ GeV.
We made use of the replica method\tcite{Forte:2002fg,Ball:2021dab} to determine the statistical uncertainty of our fit.
Further details on the fitting procedure and quality can be found in Ref.\tcite{Bacchetta:2020vty}.
Since our tree-level approximation does not depend on the gauge link, our model T-even TMDs are process-universal.
A preliminary application of our T-even gluon TMDs to the hadroproduction of pseudoscalar quarkonium states were presented in Ref.\tcite{Bacchetta:2020vty}, while the extension of our framework to the T-odd case was discussed in Refs.\tcite{Bacchetta:2021lvw,Bacchetta:2021twk,Bacchetta:2022esb,Bacchetta:2022crh,Bacchetta:2022nyv}.

With the aim of unveiling the 3D dynamics of gluons inside the proton, we investigate the following distributions which describe the 2D $\boldsymbol{\kappa}_T$-density of gluons for distinct combinations of their polarization and the nucleon spin. For an unpolarized proton, we identify the unpolarized distribution
\begin{equation}
 x {\cal F}_u^g (x, \kappa_x, \kappa_y) = x f_1^g (x, \boldsymbol{\kappa}_T^2) 
\label{eq:F_unpol}
\end{equation}
as the distribution of an unpolarized gluon with $x$ and $\boldsymbol{\kappa}_T$, whereas the Boer--Mulder distribution 
\begin{equation}
 x {\cal F}_{\leftrightarrow}^g (x, \kappa_x, \kappa_y) = \frac{x}{2} \bigg[f_1^g (x, \boldsymbol{\kappa}_T^2) - \frac{\kappa_y^2 - \kappa_x^2}{2 m_{\cal H}^2} \, h_1^{\perp g} (x, \boldsymbol{\kappa}_T^2) \bigg]
\label{eq:F_T}
\end{equation}
stands for the density of a linearly-polarized gluon in the transverse plane with $x$ and $\boldsymbol{\kappa}_T$.

Contour plots in Fig.~\ref{fig:3D_gluon_TMDs} exhibit $\boldsymbol{\kappa}_T$-patterns of $x {\cal F}$ densities in Eqs.~\eqref{eq:F_unpol} and~\eqref{eq:F_T} at $\mu_0 = 1.64$ GeV and $x=10^{-3}$ for an unpolarized proton moving towards the reader. For the sake of simplicity, results are shown for the most representative replica, namely the number 11. The color code quantifies the weight of the oscillation of each density along the $\kappa_{x,y}$ directions. The unpolarized distribution in Eq.~\eqref{eq:F_unpol} possesses a cylindrical symmetry, while the Boer--Mulders density in Eq.~\eqref{eq:F_T} has a dipolar behavior. The departure from the cylindrical symmetry is larger at low-$x$, namely in a regime where the Boer--Mulders function is expected to give a relevant contribution\tcite{Celiberto:2021zww,Bolognino:2021niq}.
To better visualize these oscillations, we present in Fig.~\ref{fig:ancillary_gluon_TMDs} ancillary, slice panels portraying the corresponding distributions at $\kappa_y = 0$ and for $x$ ranging from $10^{-1}$ down to $10^{-3}$.
We clearly observe that both our functions increase as $x$ lowers. This is in line with the well-known pattern in $x$ of collinear PDFs, from which we have extracted the small-$x$ behavior via our fit procedure.

From an analytic perspective, in our model the $f_1^g/h_1^{\perp g}$ ratio is asymptotically constant in the $x \to 0^+$ limit.
This consistently matches the prediction genuinely, coming from the linear Balitsky--Fadin--Kuraev--Lipatov (BFKL) evolution\tcite{Fadin:1975cb,Kuraev:1976ge,Kuraev:1977fs,Balitsky:1978ic}, that at small-$x$ the ``number'' of unpolarized gluons equals the linearly-polarized one, up to higher-twist effects\tcite{Metz:2011wb,Dominguez:2011br,Marquet:2016cgx,Taels:2017shj,Marquet:2017xwy,Petreska:2018cbf}).
In this way, a first intersection point between our model gluon TMDs and the high-energy dynamics has been revealed.

%-----------------------------------------
\section{Prospects}
\label{sec:conclusions}
%-----------------------------------------

We reported progresses on accessing the core of the proton via TMD gluon distributions. The presented results are relevant to explore the multidimensional structure of the proton, where the intrinsic motion of partons plays a key role to describe observables sensitive to different combinations of parton and hadron spins. In the low-$x$ domain, a key role is played by the connections between TMD and BFKL effects.
The high-energy resummation allows us to shed light on the proton structure at small-$x$ via single-forward emissions.
In particular, it gives us direct access to the \emph{unintegrated gluon distribution} (UGD) in the proton, whose evolution is regulated by the BFKL Green's function.
Pioneering analyses of the UGD were performed through the study of: deep-inelastic-scattering structure functions\tcite{Hentschinski:2012kr,Hentschinski:2013id}, light vector-meson helicity amplitudes and cross sections\tcite{Anikin:2009bf,Anikin:2011sa,Besse:2013muy,Bolognino:2018rhb,Bolognino:2018mlw,Bolognino:2019bko,Bolognino:2019pba,Celiberto:2019slj,Bolognino:2021niq,Bolognino:2021gjm,Bolognino:2022uty,Celiberto:2022fam,Bolognino:2022ndh,Cisek:2022yjj,Luszczak:2022fkf}, forward Drell--Yan\tcite{Motyka:2014lya,Brzeminski:2016lwh,Motyka:2016lta,Celiberto:2018muu} and single-forward quarkonium\tcite{Bautista:2016xnp,Garcia:2019tne,Hentschinski:2020yfm,Goncalves:2018blz,Cepila:2017nef,Guzey:2020ntc,Jenkovszky:2021sis,Flore:2020jau,ColpaniSerri:2021bla} final states.
Starting from the information on the gluon motion carried by the UGD, determinations of small-$x$ improved collinear PDFs were obtained\tcite{Ball:2017otu,Bonvini:2019wxf,Abdolmaleki:2018jln}.
The connection between the unpolarized and linearly polarized gluon TMDs, $f^g_1$ and $h^{\perp g}_1$, and the UGD was investigated in Refs.\tcite{Dominguez:2011wm,Hentschinski:2021lsh,Nefedov:2021vvy,Celiberto:2021zww}.
Future studies of the proton structure at new-generation colliding machines will benefit from the recently discovered property of \emph{natural stability} of the high-energy resummation~\cite{Celiberto:2022grc} (see also Refs.~\cite{Caporale:2012ih,Ducloue:2013hia,Ducloue:2013bva,Caporale:2013uva,Caporale:2014gpa,Ducloue:2015jba,Celiberto:2015yba,Celiberto:2015mpa,Caporale:2015uva,Mueller:2015ael,Celiberto:2016ygs,Celiberto:2016vva,Caporale:2018qnm,Celiberto:2022gji,Celiberto:2016hae,Celiberto:2016zgb,Celiberto:2017ptm,Celiberto:2017uae,Celiberto:2017ydk,Caporale:2015vya,Caporale:2015int,Caporale:2016soq,Caporale:2016vxt,Caporale:2016xku,Celiberto:2016vhn,Caporale:2016djm,Caporale:2016pqe,Chachamis:2016qct,Chachamis:2016lyi,Caporale:2016lnh,Caporale:2016zkc,Chachamis:2017vfa,Caporale:2017jqj,Bolognino:2018oth,Bolognino:2019cac,Bolognino:2019yqj,Celiberto:2020wpk,Celiberto:2020rxb,Celiberto:2022kxx,Celiberto:2020tmb,Hentschinski:2020tbi,Celiberto:2021fjf,Celiberto:2021tky,Celiberto:2021txb,Celiberto:2021xpm,Celiberto:2022fgx,Boussarie:2017oae,Celiberto:2017nyx,Bolognino:2019ouc,Bolognino:2019yls,Bolognino:2019ccd,Celiberto:2021dzy,Celiberto:2021fdp,Bolognino:2022wgl,Celiberto:2022dyf,Celiberto:2022keu,Celiberto:2022zdg,Celiberto:2022kza,Bolognino:2021mrc,Bolognino:2021hxxaux,Celiberto:2022qbh,Bolognino:2022paj,Fucilla:2022whr}).
Finally, to get a complete 3D picture of the proton, the information about small-$x$ gluon dynamics needs to be complemented by the valence-quark one at moderate- and large-$x$. A key role along this direction will be played by new data collected via future fixed-target programs, such as the JLab CLAS12~\cite{Niccolai:2011zz,Avakian:2011zz,Burkert:2012rh,Aghasyan:2013kz,Matevosyan:2015gwa,Avakian:2019drf,Proceedings:2020fyd,Hayward:2021psm} and the LHC one~\cite{Brodsky:2012vg,Lansberg:2012kf,Lansberg:2015lva,Kikola:2017hnp,Hadjidakis:2018ifr} with possible polarized targets~\cite{Aidala:2019pit,Santimaria:2021uel,Passalacqua:2022jia}.

%-----------------------------------------
\section*{Acknowledgments}
%-----------------------------------------

This work was supported by the INFN/NINPHA project.
The author thanks Alessandro Bacchetta, Marco Radici and Pieter Taels for collaboration, and the Universit\`a degli Studi di Pavia for the warm hospitality.

%\bibliographystyle{cas-model2-names}
\bibliographystyle{elsarticle-num}

%-----------------------------------------
\bibliography{bibliography}
%-----------------------------------------

\end{document}